# Detection of self-oscillations of the transport current in a doubly connected superconductor

S. I. Bondarenko, V. P. Koverya, A. V. Krevsun, N. M. Levchenko, and A. A. Shablo

*B. I. Verkin Institute for Low-Temperature Physics and Engineering, pr. Lenina 47, Kharkov 61103, Ukraine*

It has been found experimentally that when dc current is passed through a circuit consisting of two superconductors connected in parallel and reaches its critical value in one of the circuit branches the current in the branches undergoes quasi-harmonic undamped oscillations. The mechanism resulting in the appearance of the self-oscillations is discussed. The characteristics of magnetic field freezing in a circuit with self-oscillating current are examined.

## I. INTRODUCTION

Doubly connected superconductors in the form of two branches whose properties are uniform or nonuniform and which are connected in parallel are widely used in different superconducting devices. Examples are macroscopic objects such as quantum magnetic flux meters or SQUIDs and superconducting coils, which can be transferred into a short-circuited state by connecting thermal switches connected in parallel, for producing strong magnetic fields. Nanosize doubly connected and multiply connected superconducting circuits can also arise in modern granular and quasi-crystalline high-temperature superconductors.

The distribution of the dc superconducting electric current flowing through the branches of a doubly connected superconductor in a subcritical state is classic and is described in, for example, Ref. 1.

The objective of the present work is to investigate the dc current distribution in such structures when the current reaches its critical value. Before starting the present work we knew of no publications by other authors concerning this question. It should also be noted that it is highly unlikely that the critical currents and inductances of the branches made using real superconductors will be absolutely identical. For this reason the results of the present work could be of practical value.

## II. EXPERIMENTAL ARRANGEMENT

The electric circuit of the doubly connected superconductor studied here is displayed in Fig. 1.

The doubly connected circuit was made of 100 $\mu$m in diameter tantalum microwire. The inductances of the circuit branches were $L_1 = 5 \cdot 10^{-6}$ H and $L_2 \cong 10^{-8}$ H; the mutual inductance was $M = 2 \cdot 10^{-10}$ H. The branch with inductance $L_1$ comprised an 8 mm in diameter coil with $W=5$ turns. A detector for the magnetic field of the current flowing along the coil was placed inside the coil. The detector is a ferromagnetic probe (FP) with sensitivity $10^{-5}$ Oe. A predetermined relation between the current in the coil and the indications of the ferromagnetic probe was used to determine the coil current from the field measured by the detector. An electromechanical automatic plotter recorded the indications of the ferromagnetic probe. A current source provided dc current $I$ from $10^{-5}$ A to 1 A. The circuit was placed in liquid helium at temperature $T=4.2$ K. The critical current $I_{c2}$ in the branch with the lower inductance was varied from 0.7–500 mA in different circuit samples; this was accomplished by making the microwire thinner—chemical etching an approximately 1 mm long section of the wire in a 1:1 solution of hydrofluoric and nitric acids. The smallest diameter (30 $\mu$m) of the etched section was located at the center of the section.

A copper solenoid was placed on the coil to freeze the magnetic field in the circuit. A magnetic screen protected the cryostat containing the experimental circuit from the Earth's field and its fluctuations. The amplitude of the residual low-frequency fluctuations did not exceed $10^{-4}$ Oe, which corresponded to a change of the current in the coil with the ferromagnetic probe by 10 $\mu$A. A Bruel & Kjaer spectrum analyzer recorded the spectra of the self-oscillations of the current, which were measured by the ferromagnetic probe, in the frequency band 0–200 Hz.

## III. EXPERIMENTAL RESULTS AND DISCUSSION

The current–voltage characteristics (IVC) of the parallel configuration as a whole and of the second branch separately are presented in Fig. 2.

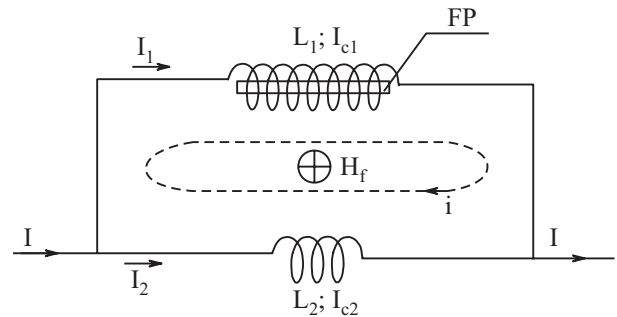

FIG. 1. Schematic of a doubly connected superconductor through which a constant current $I$ is passed; $I_1$ and $I_2$ are the currents in the branches of the superconducting circuit with inductances $L_1$ and $L_2$; $H_f$ and $i$ are the frozen magnetic field in the loop and the corresponding superconducting current; $I_{c1}$ and $I_{c2}$ are the critical currents of the branches; FP is a ferromagnetic probe for measuring the magnetic field produced by the branch current.

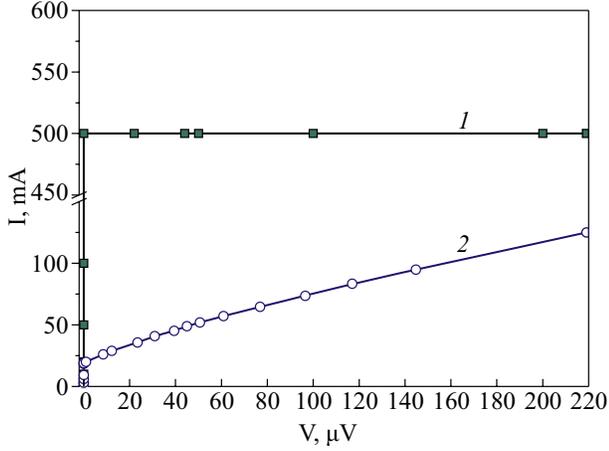

FIG. 2. Current–voltage characteristics of two tantalum superconductors connected in parallel at $T=4.2$ K (curve 1) and its branches with a "weak" section (curve 2).

As one can see, the critical current $I_c$ of the connection is 500 mA, which is determined primarily by the critical current $I_{c1}$ (about 480 mA) of the branch 1 (without the etched section). The IVC has a jumpy form and exhibits hysteresis with decreasing current (not shown in Fig. 2), which is ordinarily due to substantial overheating of the superconductor by the transport current. The critical current of the branch 2 ($I_{c2}$), measured before the branches are closed into a loop, is determined by the critical current of the "weak" etched section and equals 20 mA; its IVC does not show any hysteresis. Currents $I$ below 20 mA flow only along the branch 2 with the lower inductance. Starting at $I=I_{c2}=20$ mA a current in form of quasi-harmonic undamped self-oscillations (SO) with amplitude $\Delta I_1$ appears in the branch 1. Figure 3 displays the magnitude of the SO versus time ($t$) for different values of the constant transport current $I$.

For $I>20$ mA, together with the appearance of a dc current in the branch 1 an increase $\Delta I_1$ and a change in the form of the SO are observed. It was determined that the maximum values of $\Delta I_1$ do not exceed 1% of $I_{c2}$. The appearance of new maxima and minima in the function $\Delta I_1(t)$ as $I$ increases shows that the frequency spectrum of the SO becomes wider. Specifically, for $I \gg I_{c2}$ the spectrum of oscillations of $\Delta I_1(t)$ was recorded right up to 200 Hz. Taking the characteristic time of the oscillatory process for currents close to $I_{c2}$, $T_0 \cong 1$ s (Fig. 3), and knowing the inductance of the circuit ($L=5 \cdot 10^{-6}$ H), the resistance periodically introduced into the circuit can be estimated as $R \cong L/T_0 = 5 \cdot 10^{-6}$ Ω. For constant currents $I>20$ mA, because $I=I_1+I_2$, the corresponding current oscillations also exist in the branch 2.

Aside from studying current self-oscillations, experiments were also performed on freezing a weak magnetic field $H<0.1$ Oe, which corresponded to a circulating current (current $I$ in Fig. 1) of no higher than 20 mA in a closed circuit consisting of the indicated superconductors connected in parallel. In the presence of SO the frozen magnetic field (FMF) remained unchanged (to within 1%) for two hours.

The observed phenomena were explained as follows. A stable distribution of a constant superconducting current $I$ can be established in the branches of a doubly connected circuit in a subcritical state on the basis of the concept of minimum magnetic energy $E_m$ of a superconducting circuit with a current. This concept is preferable to that adopted in Ref. 1, since it makes it possible to explain the current distribution not only in the subcritical but also the critical state of the circuit. Since $M \ll L_1$ and $L_2$, the condition for a minimum of $E_m$ can be expressed as

$$E_m = \Phi^2/2L = (L_1I_1 - L_2I_2)^2/2L = 0, \quad (1)$$

where $\Phi$ is the magnetic flux through the loop created by the currents in the branches, $L=L_1+L_2$ is the inductance of the circuit. Thus, for $I_1<I_{c1}$ and $I_2<I_{c2}$ we have from Eq. (1)

$$I_1/I_2 = L_2/L_1, \quad (2)$$

and since $I=I_1+I_2$ we obtain

$$I_1 = (L_2/L)I, \quad (3)$$

$$I_2 = (L_1/L)I. \quad (4)$$

It follows from the relations (3) and (4) that for the experimental values of the parameters appearing in them a current $I$ of magnitude up to 20 mA indeed should flow mainly through the branch 2. The current in the branch 1 must be less than 5 $\mu$A, which falls within the limits of the sensitivity of the measuring system used. This is confirmed by the curves in Fig. 3. This feature of the threshold appearance of SO can be used to determine the critical current of the "weak" section precisely.

As the current $I$ increases to values at which the critical current is reached in one of the branches with a section that is "weak" with respect to the superconducting parameters (in our case in the branch 2), a part ($\Delta I_2$) of the current $I_2$ switches to the branch with a high critical current (into branch 1), increasing the current $I_1$ by the amount $\Delta I_1 = \Delta I_2$. As a result the magnetic flux through the circuit and the energy of the circuit increase compared with the stable subcritical state set by the relations (3) and (4), and the system strives to lower it, which should cause the indicated current to return to the initial state. Subsequently, the process repeats. The observed SO as a whole correspond to the proposed model but at the same time exhibit features which the model does not predict. These concern the form and amplitude of the current SO, which change with increasing $I$. These features can be explained by the fact that the model neglects the concrete conditions of the experiment:

— differences and slowness of the heat exchange between the "weak" section of the superconductor and the liquid helium when it is heated by the normal-electron current ($I_2 > I_{c2}$) and when it is cooled by a normal-electron current ($I_2 < I_{c2}$);
— the concrete mechanism of the resistivity of the "weak" extended (1–2 mm) section of the tantalum wire as a type-I superconductor, which, in particular, itself can consist of several successively connected microscopic regions (with different and higher than 20 mA critical currents), increasing the initial resistivity of the section as the current $I$ increases;
— changes of the ratio $I/R$, where $R$ is the time-varying and current-dependent resistivity of the "weak" section of the circuit.

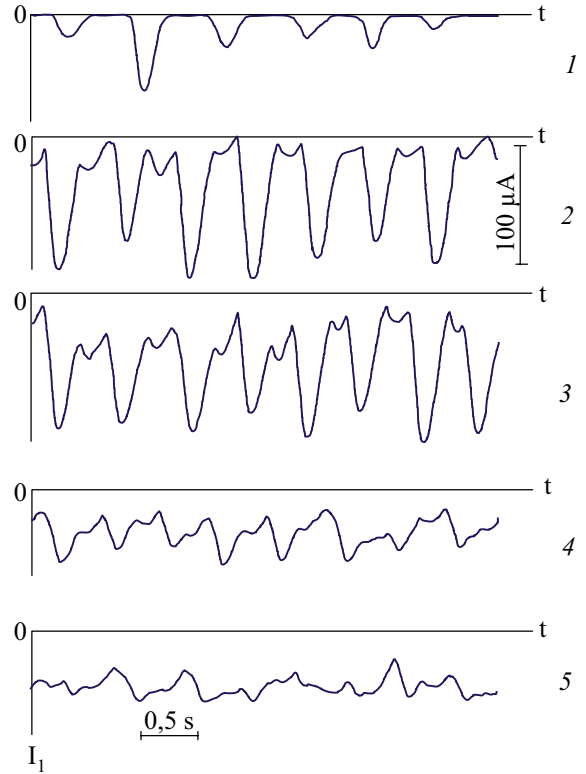

FIG. 3. Trace of the self-oscillations of the current $I_1$ flowing in one of the branches of the superconducting circuit (branch 1) for different values of the constant transport current $I$, mA: 20 (1), 20.3 (2), 21.0 (3), 21.5 (4), and 22.5 (5).

In subsequent investigations we intend to elucidate the effect of these conditions on the SO.

The fact that the FMF remains in the circuit with SO for a long time shows that the dissipation of the magnetic energy stored in the circuit as a result of the freezing is very small.

The estimated time during which the weak frozen field $H_f = 0.1$ Oe remains in the circuit with energy dissipation on the resistance $R \cong 10^{-6}\ \Omega$ as obtained by comparing the magnetic energy of the frozen flux $Li^2/2$ and the thermal energy $I^2R\Delta t$ with $I = 20$ mA showed that the FMF should vanish completely in a time ($\Delta t$) equal to several seconds, which is at variance with the experiment.[2] This is probably due to the approximate nature of the estimate of the resistance of the "weak" section of the circuit and requires further refinements.

In conclusion it should be noted that the experimental results and explanation of the quasi-harmonic SO as presented above make it possible to understand the reason for the appearance of the self-oscillations of the voltage on a resistive dc SQUID in a superconducting ring, which were previously observed by one of the present authors.[3]

**CONCLUSIONS**

The quasi-harmonic self-oscillations of the transport current observed in a doubly connected superconductor can be explained on the basis of the following ideas: the current reaching the critical state in the branch with the lowest critical value, part of the current switching into the branch with the higher critical current, and the system striving to restore subsequently a state with the lowest magnetic energy by the means of the switched current returning to the initial position. It was determined that the switched part of the current in the experimental superconducting circuit (with the branch inductances in the ratio 1:500) does not exceed 1% of the critical value of the branch with the lowest inductance.

The form and amplitude of the self-oscillations depend on the amount by which the transport current exceeds its critical value in one of the branches of the doubly connected superconductor and can be related with the particulars of the formation of the resistance of the section of the branch in which the critical current is reached.

The long time during which the frozen magnetic field and the corresponding circular current (not exceeding the lowest critical current of the branches of the loop) last in a doubly connected type-I superconductor with current SO present attests to the extremely weak dissipation of the magnetic energy of the field frozen in the circuit. This requires further study.

The appearance of self-oscillations with a transport current flowing through a doubly connected superconducting circuit equal to the threshold value at which an extremely low resistance appears in one of its branches can be used as an alternative method of determining the critical constant current of different superconducting structures, first and foremost, structures containing a small normal resistance (of the order of $10^{-6}\ \Omega$).